\documentstyle[aps,prl,multicol,epsfig]{revtex}

\begin{document}

\title{Statistical properties of phases and delay times of the
one-dimensional Anderson model with one open channel}
\author{A. Ossipov, Tsampikos Kottos, and T. Geisel
\\
Max-Planck-Institut f\"ur Str\"omungsforschung und Fakult\"at Physik
der Universit\"at G\"ottingen,\\
Bunsenstra\ss e 10, D-37073 G\"ottingen, Germany} 
\maketitle

\begin{abstract} 
We study the distribution of phases and of Wigner delay times for a one-dimensional 
Anderson model with one open channel. Our approach, based 
on classical Hamiltonian maps, allows us an analytical treatment. We find that the 
distribution of phases depends drastically on the parameter $\sigma_A = \sigma/sin k$ 
where $\sigma^2$ is the variance of the disorder distribution and $k$ the wavevector. 
It undergoes a transition from uniformity to singular behaviour as $\sigma_A$ 
increases. The distribution of delay times shows universal power law tails $~ 1/\tau^2$, 
while the short time behaviour is $\sigma_A$- dependent.
\\

\end{abstract}

\begin{multicols}{2}

Quantum mechanical scattering by chaotic or disordered systems, has been a 
subject of a rather intensive research activity during the last ten years. 
This interest was motivated by various areas of physics, ranging from nuclear 
\cite {nuclear}, atomic \cite {atomic} and molecular \cite {molecular} physics, 
to mesoscopics \cite {mesoscopics} and classical wave scattering \cite{S89}. 
The most fundamental object characterizing the process of quantum scattering 
is the unitary $S$-matrix relating the amplitudes of incoming waves to the 
amplitudes of outgoing waves. At present, there are two complementary theoretical 
tools employed to calculate statistical properties of the $S-$matrix, namely 
the semiclassical and the stochastic approach. The starting point of the first 
is a representation of the $S-$ matrix elements in terms of a sum of classical 
orbits \cite{S89,KS99} while the later exploits the similarity with ensembles 
of Random Matrices \cite{FS97}. Thus, for chaotic systems, many results 
are known. Namely, the distribution of the $S-$matrix was found to be described 
by the so-called Poisson Kernel \cite{KS99,FS97}. In the same framework, the 
Wigner delay time statistics has been studied intensively. This quantity captures 
the time- dependent aspects of quantum scattering. It is related to the time 
spent in the interaction region by a wavepacket of energy peaked at $E$. The delay 
time is not self-averaging and one must have its full probability distribution. 
For the one-channel case this was found in \cite{FS97} while the case with 
generally $M$ open channels was presented in \cite{BFB97}. 

In spite of the wealth of new results on chaotic scattering, very little is known 
for the scattering from systems being in the localized regime \cite{BG93}, even 
for the simplest case of one dimensional (1D) disordered systems where Anderson 
localization dominates \cite{JGJ97}. For the latter case and for only one channel, 
there are some analytical results about the distribution of phases of the $S-$ 
matrix and of delay times. They were first derived with the help of the invariant 
embedding method \cite{JVK89} and more recently with the use of exponential 
functionals of Brownian motion \cite{TC99}. In all cases, however, these results 
are restricted to continuous systems and to weak disorder.

In this letter, we surmount these limitations by using a different approach, which 
transforms the initial 1D tight-binding Anderson model into a classical linear oscillator with a 
parametric perturbation given in the form of periodic $\delta$-kicks \cite{IKT95,IRT98}. 
The amplitudes of these kicks are defined by the ratio $A_n= V_n/sin k$, where $V_n$ 
is the on-site random potential with variance $\sigma^2$ and $k$ is the wavevector. Based on this approach we derive simple iteration relations 
for the phases of the $S$-matrix and the delay times which allow us to analyze their 
distributions. We find that the kick strength $\sigma_A = \sigma/sin k$-- and not $\sigma$ itself \cite{JGJ97,JVK89,TC99}-- is a new significant parameter that 
controls the form of these distributions . We challenge 
-- even for the weak disorder regime $\sigma\ll 1$ -- the one-parameter scaling based on 
the so-called random phase hypothesis \cite{AALR79}, and conclude, in agreement with 
\cite{DLA99}, for a new criterion for phase randomization which is $\sigma_A \ll 1$. 
In this limit, we derive the distribution of delay times analytically. It is found 
to be insensitive to the specific type of the random potential, but may 
depend on its variance. On the other hand, in the opposite limit $\sigma_A\gg 1$, 
it becomes disorder dependent. The tail however follows the same power 
law $~1/\tau^2$, as in the $\sigma_A\ll 1$ limit. 

We consider a 1D disordered sample of length $L$ with one semi-infinite perfect 
lead attached on the left side. The system is described by the tight-binding equation:
\begin{equation}
\label {tight-binding}
\psi_{n+1} + \psi_{n-1} = (E-V_n)\psi_{n};\,\,\,\,E=2cos k
\end{equation}
where $\psi_n$ is the wavefunction amplitude at site $n$. 
We assume that for $0\leq n\leq L$, $V_n$ is random 
delta-correlated given by a distribution ${\cal P}_V$ with mean zero and 
variance $\sigma^2$ . For $n < 0$, $V_n=0$ and we impose Dirichlet boundary 
conditions at the edge $\psi_{L+1} = 0$. Therefore, for $n\leq 0$, scattering 
states of the form $\psi_n  = {\rm e}^{ikn} + r{\rm e}^{-ikn}$ represent the 
superposition of an incoming and a reflected plane wave. Since there is only 
backscattering, the reflection coefficient $r(E)\equiv e^{i\Phi(E)}$ is of 
unit modulus and the total information about the scattering is contained in 
the phase $\Phi(E)$. The Wigner delay time is given by $\tau(E) \equiv 
\frac{d\Phi(E)} {dE}$. Our aim is to find the probability distribution of the 
phases ${\cal P}_{\Phi}(\Phi)$ and of delay times ${\cal P}_{\tau}(\tau)$.

Eqn.~ (\ref{tight-binding}) can be written equivalently in a form
\begin{eqnarray}
\label{eqM}
{\left (
\begin {array} {c}
\psi_{n+1}  \\ \psi_{n}
 \end {array}
\right )} =
 M_n
\left (
\begin {array} {c}
\psi_{n}  \\ \psi_{n-1}
 \end {array}
\right );
M_n =
\left (
\begin{array}{cc}
E - V_n  &  -1 \\
1 & 0
\end{array}
\right ).
\end{eqnarray}
Consequently we have the following relation for the total transfer 
matrix $P = \prod^L_{n=0} M_n $
\begin{eqnarray}
\label{piter}
{\left (
\begin {array} {c}
0  \\ \psi_{L}
 \end {array}
\right )} =
\left (
\begin{array}{cc}
P_{11} &  P_{12} \\
P_{21} & P_{22}
\end{array}
\right )
\left (
\begin {array} {c}
1 + r  \\{\rm e}^{-ik} + r{\rm e}^{ik}
 \end {array}
\right ).
\end{eqnarray}
Solving Eqn.~(\ref{piter}) for $r$ we get
$
r = - \frac{P_{11}+P_{12}{\rm e}^{-ik}}
{P_{11}+P_{12}{\rm e}^{ik}}
= {\rm e}^{2i\phi}
$
where $\phi = \Phi/ 2$ is now given by
\begin{eqnarray}
\label{phi}
tan \phi = \left(\frac{P_{11}+P_{12}cosk}{P_{12}sink}\right)
\end{eqnarray}
As was indicated in \cite{IKT95}, one can rewrite (\ref{eqM}) in 
the form of a two-dimensional Hamiltonian map $Q_n$ 
\begin{eqnarray}
\label{Hamiltonianmap}
{\left (
\begin {array} {c}
x_{n+1}  \\ p_{n+1}
 \end {array}
\right )} =Q_n
\left (
\begin {array} {c}
x_n  \\p_n
 \end {array}
\right );\,
Q_n = 
\left (
\begin{array}{cc}
cosk-A_nsink &  sink \\
-A_ncosk-sink & cosk
\end{array}
\right ).
\end{eqnarray}
Here, $(x_n,p_n)$ play the role of a position and momentum of a parametric
linear oscillator subjected to periodic kicks of strength $A_n = \frac{V_n}{sink}$
and period $T=1$. Between two successive kicks, there is a free rotation in the phase
space which is determined by the eigenenergy $E$ of our initial equation 
(\ref{tight-binding}). In such a representation, amplitudes $\psi_n$ of a specific 
eigenstate correspond to positions of the oscillator at times $t_n=n$. $Q_n$ 
is related to the transfer matrix $M_n$ through a similarity transformation
\cite{IKT95}:
\begin{equation}
Q_n=RM_nR^{-1};\,\,\,
R=
\left (
\begin{array}{cc}
1 &  0 \\
cosk/sink & -1/sink
\end{array}
\right ) .
\end{equation}
As a result, the total transfer matrix $P$ is related to the map $F=\prod^L_{n=0} Q_n$ 
through the similarity transformation $P=R^{-1}FR$. Using this, together with 
(\ref{phi}) we write $\phi$ in terms of our Hamiltonian map $F$ as:
\begin{equation}
\label{phimap}
tan \phi = \left( \frac{F_{11}}{-F_{12}}\right)
\end{equation}
We can give a geometrical interpretation for (\ref{phimap}). Consider the
time evolution of the vector $ v(t=0) = ( 0,1)^T$ under the inverse map $F^{-1}$. 
For time $t_n = L$ we have
\begin{eqnarray}
v(t=L) = F^{-1}
{\left (
\begin {array} {c}
0  \\ 1
 \end {array}
\right )}=
{\left (
\begin {array} {c}
-F_{12}  \\ F_{11}
 \end {array}
\right )}.
\end{eqnarray}
Then $\phi$ is exactly the angle between the vector $v(t=L)$ and the x-axis. It is convenient 
therefore to pass to polar coordinates $(r_n,\theta_n)$, using the transformation $x=rcos\theta$ 
and $p=rsin\theta$. Then (\ref{Hamiltonianmap}) is written in the form \cite{IKT95}:
\begin{eqnarray}
r_{n+1}^2 &=& r_n^2 D_n\,; D_n^2=(1+A_n^2cos^2\theta_n-A_nsin2\theta_n)\nonumber \\
cos\theta_{n+1}&=& D_n^{-1}[cos(\theta_n+k)+A_ncos\theta_n sink]\nonumber \\
sin\theta_{n+1}&=& D_n^{-1}[sin(\theta_n+k)-A_ncos\theta_n cosk].
\label{polar}
\end{eqnarray}

The relation between $\phi_{n+1}$ and $\phi_n$ is found after inversion of (\ref{polar})
and has the form \cite{felix}:
\begin{equation}
\label{revphi}
tan(\phi_{n+1}) = tan(\phi_n-k)+A_n
\end{equation}

From (\ref{revphi}), we get the following equation for the stationary ($n\rightarrow \infty$) 
distribution of phases ${\cal P}_{\phi}(\phi)$:
\begin{equation}
\label{disphi1}
\frac{{\cal P}_{\phi}(atan y)}{1+y^2} = \int ds \frac{{\cal P}_{\phi}(atan(y-s)+k)}
{1+(y-s)^2}{\cal P}_{A}(s).
\end{equation}

In the $\sigma_A\ll 1$ limit, we can approximate ${\cal P}_{A}(s)$ by a 
$\delta-$function. Then from (\ref{disphi1}) we have for the ${\cal P}(\Phi)$
\begin{equation}
\label{disphi2}
{\cal P}_{\Phi}(\Phi) = {\cal P}_{\Phi} (\Phi+k) 
\end{equation}
which for $k$ equals to irrational multiples of $\pi$, leads to a uniform distribution 
\begin{equation}
\label{weakphi}
{\cal P}_{\Phi}(\Phi) =\frac 1{2\pi};\,\,\,k \notin  Q 
\end{equation}
in agreement with previous numerical results \cite{JGJ97}. We understand (\ref{weakphi}) 
in the following way: For $\sigma_A\ll 1$, the particle travels long distances 
inside the sample and undergoes many scattering events which leads to randomization of 
the phase. We should stress that for $k$ equals to rational values of $\pi$, the 
distribution is non-uniform. For example, for $k=\pi/2$ we obtain \cite{IRT98,felix,OKG99}
\begin{equation}
\label{disphi3}
{\cal P}_{\Phi}(\Phi) =\left( 2{\cal K}(\frac 1{\sqrt 2}) 
{\sqrt{ 3+cos(2\Phi)}}\right)^{-1};\,\,\,\,\,\,\,\,k=\frac{\pi}2
\end{equation}
where ${\cal K}$ is the complete elliptic integral of the first kind. A similar anomaly at 
the band center governs the distribution of the angles $\theta$ of the direct map (\ref{polar}) 
\cite{IRT98}. The above expressions (\ref{weakphi},\ref{disphi3}) describe well the
corresponding numerical results presented in Figs.~1a,b.

In the $\sigma_A \gg1 $ limit, we distinguish between two parts of the spectrum. 
Namely, $k$ near the band edges (i.e. $k\simeq 0,\pi$) and far away from them. For the
latter case and for $V_n$ distributed uniformly between $[-\frac V2;\frac V2]$ one can derive 
an analytical expression for ${\cal P}_{\Phi}$ by rewriting (\ref{revphi}) in terms
of the probability distributions ${\cal P}_{tan(\phi)}$. We get (in first order in $1/\sigma_A$) 
the uniform distribution \cite{OKG99}
\begin{equation}
\label{disphi5}
{\cal P}_{tan(\phi)}(y) = ({\sqrt 12}\sigma_A)^{-1}
\Theta\left( {\sqrt 3} \sigma_A  - |y-cot k|\right) .
\end{equation}
where $\Theta(x)$ is the Heavyside function. The distribution of phases 
${\cal P}_{\Phi}(\Phi)$ can be derived easily from (\ref{disphi5}):
\begin{equation}
\label{strongphi}
{\cal P}_{\Phi}(\Phi) = \frac {0.5}{{\sqrt 3} \sigma_A} \frac {\Theta\left({\sqrt 3}\sigma_ A-
\left|tan(\frac{\Phi}2) -cot k\right| \right)} {cos^2(\frac{\Phi} 2 )}.
\end{equation}
In contrast with the $\sigma_A\ll 1$ limit (\ref{weakphi}), here the distribution of phases 
(\ref{strongphi}) is highly non-uniform. By increasing the disorder strength $\sigma$, two peaks 
appear in the neighbourhood of $\Phi = \pi$ 
(see Fig.~1c) while a gap is created between them. As $\sigma_A$ increases further, the two peaks 
move closer to one another. Now the scattering is so strong that most of the particles are 
reflected back from near the surface, thus being scattered only from a few sites. In the 
limit $\sigma_A\rightarrow \infty$, the distribution becomes a $\delta$ function centered at 
$\Phi = \pi$. (\ref{strongphi}) is in very good agreement with our numerical data presented 
in Fig.~1c. The behavior near $\Phi = \pi$ is more subtle than the one given 
above; nevertheless, (\ref{strongphi}) gives the correct scale on which the distribution 
vanishes near $\Phi = \pi$. Distribution (\ref{strongphi}) agrees quite nicely also for 
other disorder potentials \cite{OKG99}. 

Near the band edge ($k\approx 0,\pi$), our detail numerical analysis showed that 
${\cal P}_{\Phi}(\Phi)$ is again highly non- uniform. Namely the distribution becomes narrower 
and centered (although with a slight asymmetry) at $\Phi = \pi$ . We notice that 
such a choice of the parameters, can be realized even for weak disorder $\sigma \ll 1$. In Fig.1d 
we present a representative case corresponding to $k=10^{-3}\sqrt{\pi}$ and $\sigma= 0.0577$.
\vspace*{-1.2cm}
\begin{figure}
\hspace*{-1cm}\epsfig{figure=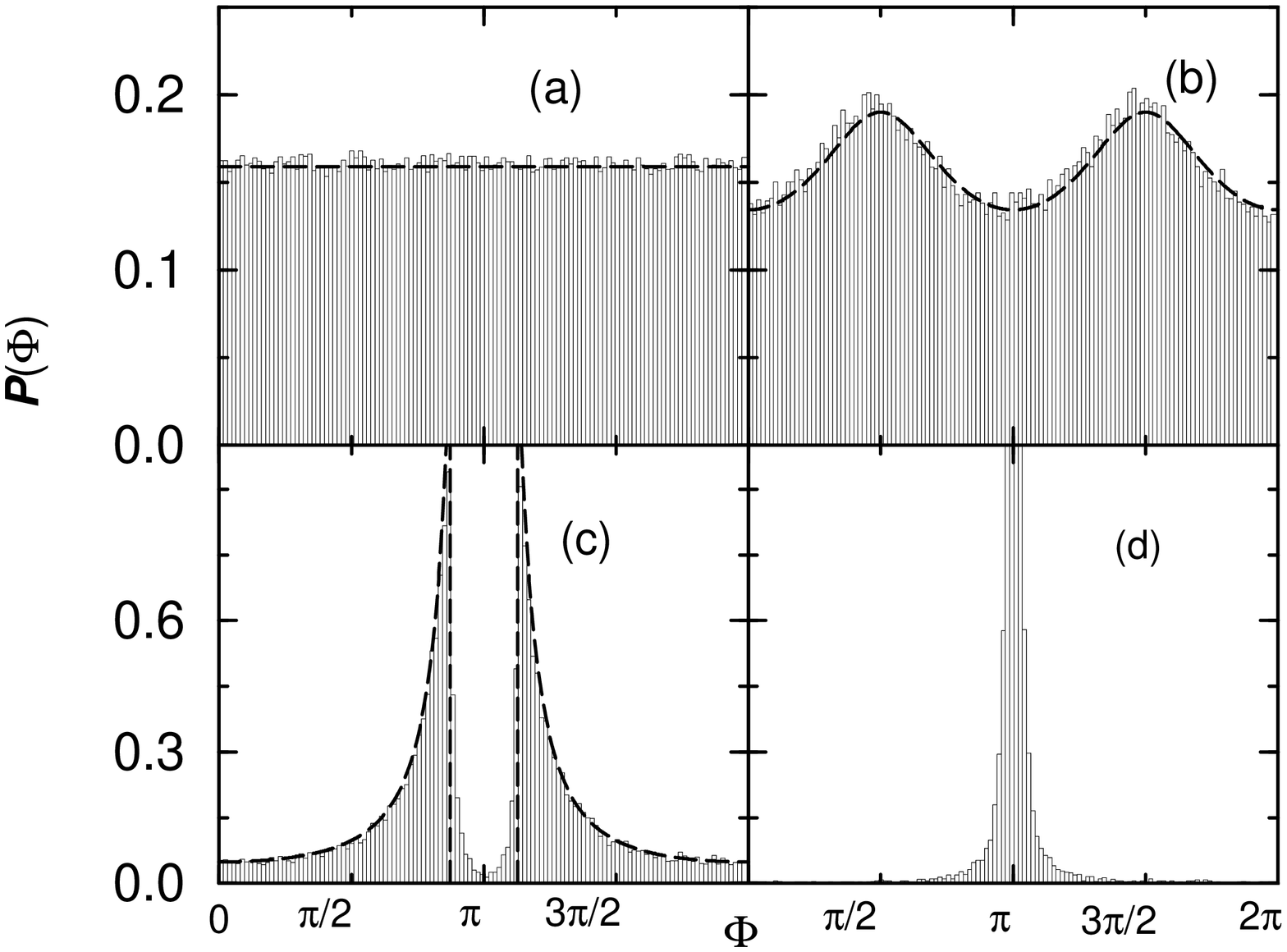,height=10cm,width=7cm,angle=270}
\noindent
{\footnotesize {\bf FIG. 1.}
Distribution of the phases ${\cal P}(\Phi)$ for various disordered strengths and energies. 
The on-site potential is uniformly distributed between $[-\frac V2;\frac V2]$. {\bf (a)} 
$V=0.2$ and $k=\sqrt \pi$. {\bf (b)} $V=0.2$ but now for $k=\pi/2$ (band center); 
{\bf (c)} $V=10 $ and $k=\sqrt \pi$. Our numerical data (histogram) are in perfect agreement 
with the analytical predictions (dashed lines). {\bf (d)} $V=0.2$ and $k=10^{-3}\sqrt{\pi}$}. 
\end{figure}

Let us now turn to the analysis of delay times. From (\ref{revphi}) we get the following
iteration relation for $\tau_n$
\begin{eqnarray}
\label{tau}
\tau_{n+1} &=& G_n^{-1} \left(\tau_n +\frac 1{sin k}\right) + 
\frac {A_n} {1+\left(tan(\phi_n-k)+A_n\right)^2} \frac{cot k}{sin k}\nonumber\\
G_n &=& 1+A_n sin\left(2(\phi_n-k)\right)+A_n^2cos^2(\phi_n-k)
\end{eqnarray}
which proves to be very convenient for numerical calculations since it anticipates 
the numerical differentiation which is a rather unstable operation. 

For $\sigma_A\ll 1$ and $k$ equals to irrational multiples of $\pi$, we obtain an 
analytical expression for ${\cal P}_{\tau}(\tau)$. To this end, we first derive from (\ref{revphi}) 
and (\ref{tau}) a stochastic equation for the phases $\phi$ and rescale delay times ${\tilde \tau} 
= \sigma^2 \tau$ (up to $\sigma_A^2$), respectively \cite{OKG99}:
\begin{eqnarray}
\label{lang1}
\frac{d\phi}{dt} &\simeq & -k - \sigma_A^2 sin(\phi-k) cos^3(\phi-k) +
cos^2(\phi-k) A\,\,\,\nonumber\\
\frac{{d\tilde \tau}}{dt} &\simeq & -\sigma_A^2 \left( {\tilde \tau}(cos^2(\phi-k)
-sin^22(\phi-k)) -sin k\right) \nonumber\\
& &- {\tilde \tau} sin (\phi-k) A . 
\end{eqnarray}
Using the van Kampen lemma, from (\ref{lang1}) we obtain the Fokker-Planck equation for the
joint probability distribution of $\phi$ and ${\tilde \tau}$. Using the fact that
$\phi$ follows the uniform distribution (see Eqn.~(\ref{weakphi})) and assuming that
the variables $\phi$ and ${\tilde \tau}$ are statistically independent, we obtain
the Fokker-Planck equation for ${\cal P}_{\tilde \tau}({\tilde \tau})$
\begin{eqnarray}
\label{fptau}
\frac {\partial {\cal P}_{\tilde \tau}({\tilde \tau},t)}{\partial t} &=&
\frac{\sigma_A^2}{4}\left[ \frac {\partial}{\partial {\tilde \tau}}
\left( ({\tilde \tau}-4sin k){\cal P}_{\tilde \tau}({\tilde \tau},t)\right) \right. \nonumber\\
& &\left. +\frac {\partial}{\partial {\tilde \tau}} \left({\tilde \tau} 
\frac {\partial}{\partial {\tilde \tau}}({\tilde \tau} 
{\cal P}_{\tilde \tau}({\tilde \tau},t))\right)\right]. 
\end{eqnarray}
The resulting stationary distribution is obtained by setting $\frac{d{\cal P}_{\tilde \tau}}
{dt} = 0$ and has the form
\begin{equation}
\label{weaktau}
{\cal P }(\tau) = \frac {l_{\infty}}{v\tau^2} exp(-l_{\infty}/v\tau);\,\,\,\,\sigma_A \ll 1
\end{equation}
where $l_{\infty} = 2 (4-E^2)/\sigma^2$ is the localization length and $v= |\partial E/ 
\partial k|$ is the group velocity. We note that (\ref{weaktau}) is independent of the nature 
of the disorder; it only depends on its second moment 
through the localization length. (\ref{weaktau}) takes its maximum value at $\tau_{max} = 
0.5 l_{\infty}/v$, indicating that the most probable trajectory that an electron travels 
(forth and back) before it scatters outside the sample is the mean free path $l_M = l_{\infty}/4$. 
Our numerical results (see Fig.~2a) are in perfect agreement with (\ref{weaktau}). An 
expression similar to (\ref{weaktau}) was obtained in \cite{JVK89} and more recently in 
\cite{TC99} using totally different approaches. It is important to stress here that the latter 
analytical results refer to continuous disordered models.

Finally we discuss the distribution of delay times ${\cal P }(\tau)$ for $\sigma_A\gg 1$. 
In this limit, we were not able to derive any analytical expression. Our iteration relation 
(\ref{tau}) however, has proven very efficient for numerical investigations. In Fig.~2b we 
show the distribution of the delay times for a uniform and a Gaussian on-site potential 
distribution with the same variance $\sigma^2 = 10$. It is clear that the short time 
distribution differs considerably in the two cases and also with respect to 
the theoretical prediction (\ref{weaktau}). On the other hand, the distribution of large 
delay times, reflecting the times for which the wave penetrates deeper into the sample shows the 
same $1/\tau^2$ behavior independent of the form of the disorder potential. It is this 
part of the reflected wave, and not the prompt part that is expected to behave in a universal 
way \cite{TC99}. The asymptotics of the distribution for large $\tau$ is presented in 
the inset of Fig.~2b, where we plot the integrated distribution $I(\gamma) = \int_0^{\gamma} 
P(\gamma') d\gamma'$ of the inverse delay time $\gamma = 1/\tau$ in $log-log$
scale. To this end, we calculated $~10^7$ delay times using the iteration relation (\ref{tau}). 
In both cases (Gaussian and uniform distribution) presented in the inset of Fig.2b we 
collected at least $10^4$ delay times that were larger than $\tau > 150$. Our numerical 
data clearly show that $I(\gamma) \sim \gamma$ for $\gamma \ll 1$. Thus ${\cal P }(\tau) 
\sim 1/\tau^2$ in agreement with the above mentioned universal behavior \cite{JGJ97,TC99}.
\vspace*{-1.2cm}
\begin{figure}
\hspace*{-1cm}\epsfig{figure=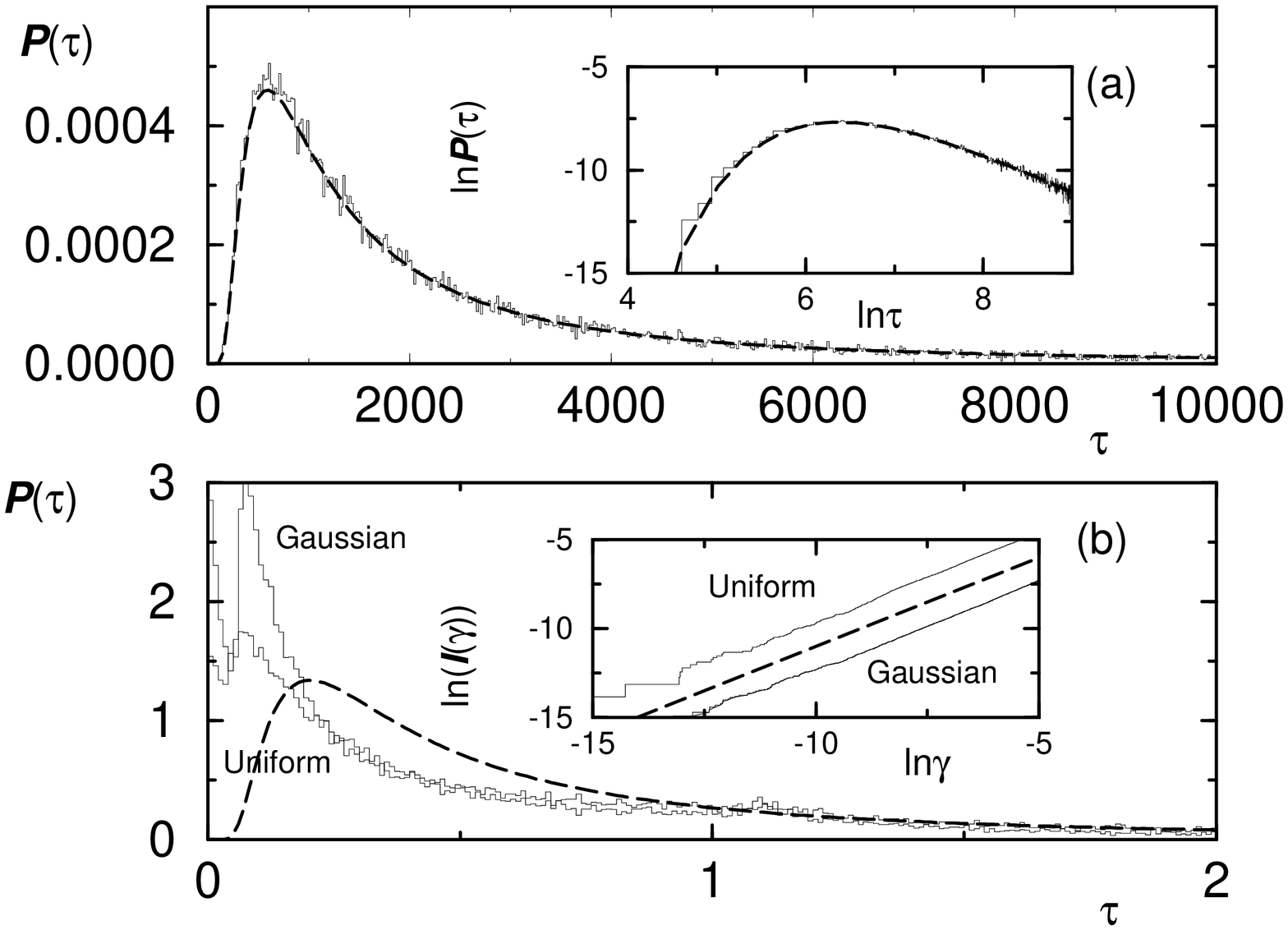,height=10cm,width=7cm,angle=270}
\noindent
{\footnotesize {\bf FIG. 2.}
{\bf (a)} Distribution of the delay times ${\cal P}(\tau)$ for on-site 
potential, uniformly distributed between $[-0.1 ;0.1]$ and wavenumber $k=\sqrt \pi$. 
The dashed line corresponds to (\ref{weaktau}).
{\bf (b)} Distribution of the delay times ${\cal P}(\tau)$ for uniform and
Gaussian ${\cal P}_V$. In both cases the variance is
$\sigma^2 = 10$ and the wavenumber is $k=\sqrt \pi$. The dashed line 
corresponds to (\ref{weaktau}). The inset shows the integrated distributions of 
inverse delay time $\gamma = 1/\tau$. The dashed line has slope $1$ and is
drawn to guide the eye.}
\end{figure}

To summarize, we analyzed the distributions of phases ${\cal P}(\Phi)$ and delay times 
${\cal P}(\tau)$ for a 1D Anderson model with one channel in the localized regime. 
For ${\cal P} (\Phi)$ we found that it undergoes a transition from uniform ($\sigma_A< 1$) 
to non-uniform behaviour ($\sigma_A>1$), where $\sigma_A = \sigma/sin k$. The latter 
limit can be achieved either by 
decreasing the disorder strength $\sigma$, or by taking $k\rightarrow 0,\pi$. This has 
further implications for the so-called one-parameter scaling hypothesis. The scaling theory 
of disordered systems \cite{AALR79} assumes 
that ${\cal P}(\Phi)$ is always uniform in the weak disorder limit $\sigma\ll 1$ leading to the
one parameter scaling. Violation of the latter is always associated with an increase of $\sigma$. 
Our analytical and numerical results showed however that a strong violation of the phase 
uniformity occurs even in the case of weak disorder $\sigma\ll 1$, in an apparent contradiction with 
the existing picture. In this limit, the states near the band edge $k\approx 0,\pi$ never 
obey the single-scaling hypothesis since $\sigma_A\gg 1$. Thus the spectrum of the system is 
divided into two groups with different scaling properties, which coexist at 
the same strength of disorder \cite{DLA99}. For ${\cal P}(\tau)$ we found that only the 
short time behaviour is affected as we are increasing $\sigma_A$. For $\tau\rightarrow \infty$, 
${\cal P }(\tau)\sim 1/\tau^2$ independent of the value of $\sigma_A$. This leads to a 
logarithmic divergence of the average value of $\tau$ indicating the possibility of the 
particle traversing the infinite sample before being totally reflected. As was indicated
in \cite{JVK89,TC99} this is another manifestation of the fact that in the localized regime 
the conductance shows lognormal distribution due to the presence of Azbel resonances. 

We acknowledge useful discussions with F. Izrailev and I. Guarneri. (T.K) thanks
U. Smilansky for initiating his interest in quantum scattering.

\end{multicols}

\begin{thebibliography}{99}

\vspace*{-1.2cm}
\bibitem{nuclear} I. Rotter, Rep. Prog. Phys. {\bf 54}, 635 (1991).

\bibitem{atomic} {\it Atomic Spectra and Collisions in External Fields},
M. H. Nayfeh et. al., eds. (Plenum, New York), Vol. 2 (1989).

\bibitem{molecular} P. Gaspard, in {\it ``Quantum Chaos", Proceedings of
E. Fermi Summer School 1991}, G. Casati et. al., eds. (North-Holland)
307.

\bibitem{mesoscopics} D. Stone in {\it Proc.~1994 Les Houches Summer
School on Mesoscopic Quantum Physics}, E.~Akkermans et.al., eds.
(North-Holland) 373-433.

\bibitem{S89}  U. Smilansky, in {\it Les Houches Summer School on Chaos
and Quantum Physics}, M.-J. Giannoni et.al., eds. (North-Holland) 371-441 (1989).

\bibitem{KS99} Tsampikos Kottos and U. Smilansky, chao-dyn/9906008. 

\bibitem{FS97} Y. Fyodorov, H-J Sommers, J. Math. Phys. {\bf 38}
1918 (1997); P. Mello in {\it Les Houches Summer School on Chaos and
Quantum Physics}, E. Akkermans et.al., eds (North-Holland) 437-491 (1994).

\bibitem{BFB97} P. W. Brouwer, K. M. Frahm, C. W. Beenakker, Phys. Rev. Lett.
{\bf 78}, 4737 (1997).

\bibitem{BG93} F. Borgonovi, I. Guarneri, Phys. Rev. E {\bf 48}, R2347 (1993).

\bibitem{JGJ97} A. D. Stone, D. C. Allan, J. D. Joanopoulos, Phys. Rev B 
{\bf 27}, 836 (1983); Sandeep K. Joshi, Abhijit Kar Gupta and A. M. Jayannavar, 
cond-mat/9712251. 

\bibitem{JVK89} A. M. Jayannavar, G. V Vijayagovindan, N. Kumar, Z. Phys. B
{\bf 75}, 77 (1989); J. Heinrichs, J. Phys.: Condens. Matter {\bf 2} 1559 (1990);
S. Anantha Ramakrishna and N. Kumar, cond-mat/9906098.

\bibitem{TC99} Christophe Texier and Alain Comtet, Phys. Rev. Lett., {\bf 82}, 
4220 (1999); C. J. Bolton-Heaton, C. J. Lambert, V. I. Falko, V. Prigodin,
and A. J. Epstein, cond-mat/9902335 (1999).


\bibitem{IKT95} Tsampikos Kottos, G.~P.~Tsironis and F.~M.~Izrailev, 
J. Phys.:Condens. Matter, {\bf 9}, 1777 (1997);  F. M. Izrailev, T. Kottos, 
G. P. Tsironis, Phys. Rev. B {\bf 52}, 3274 (1995) 

\bibitem{IRT98}F. M. Izrailev, S. Ruffo, L. Tessieri, Journ. Phys. A: Math. Gen. 
{\bf 31}, 5263 (1998).

\bibitem{AALR79}E. Abrahams, P. W. Anderson, D. C. Licciardello, T. V. Ramakrishnan,
Phys. Rev. Lett. {bf 42}, 673 (1979).

\bibitem{DLA99}L. I. Deych, A. A. Lisyansky, B. L. Altshuler, cond-mat/9909155 (1999).

\bibitem{felix} The angles $\theta$ and $\phi$ of the maps (\ref{polar}) and 
(\ref{revphi}) respectively are related by $\phi_n=\theta_n-k, \forall n$. We thank 
F. Izrailev for this comment.


\bibitem{OKG99} A. Ossipov, T. Kottos, T. Geisel, in preparation (1999).




\end{thebibliography}
\end{document}